\documentclass[twocolumn,aps,showpacs,superscriptaddress]{revtex4-1}
\usepackage{amsmath}
\usepackage{graphicx}
\usepackage{slashed}
\RequirePackage[pagewise,mathlines]{lineno}

\newcommand{\macro}[1]{\texttt{\textbackslash#1}}

\begin{document}
\title{Acoplanarity of QED pairs accompanied by nuclear dissociation in ultra-peripheral heavy ion collisions}%
\author{James Daniel Brandenburg}\affiliation{Shandong University, Jinan, China}\affiliation{Brookhaven National Laboratory, New York, USA}
\author{Wei Li}\affiliation{Rice University, Houston, USA}

\author{Lijuan Ruan}\affiliation{Brookhaven National Laboratory, New York, USA}
\author{Zebo Tang}\affiliation{University of Science and Technology of China, Hefei, China}
\author{Zhangbu Xu}\affiliation{Brookhaven National Laboratory, New York, USA}\affiliation{Shandong University, Jinan, China}
\author{Shuai Yang}\affiliation{Rice University, Houston, USA}
\author{Wangmei Zha}\email{first@ustc.edu.cn}\affiliation{University of Science and Technology of China, Hefei, China}
\date{\today}%
\begin{abstract}
This paper investigates the transverse momentum broadening effect for electromagnetic production of dileptons in ultra-peripheral heavy ion collisions accompanied by nuclear dissociation. The electromagnetic dissociation probability of nuclei for different neutron multiplicities is estimated, which could serve as a centrality definition (i.e. impact parameter estimate) in ultra-peripheral collisions. In the framework of lowest-order QED, the acoplanarity of dilepton pairs is calculated for different neutron emission scenarios in ultra-peripheral collisions, indicating significant impact-parameter dependence. The verification of impact-parameter dependence is crucially important to understand the broadening effect observed in hadronic heavy-ion collisions.
\end{abstract}
\maketitle

Collisions of heavy nuclei at ultra-relativistic energies with nuclear overlap are usually performed to study the properties of the Quark-Gluon Plasma (QGP) --- a deconfined state of partonic matter~\cite{PBM_QGP}. The dileptons from violent hadronic interactions have been proposed as ``penetrating probes'' of the hot and dense medium~\cite{SHURYAK198071}, because they are created during the whole evolution and are not sensitive to the violent strong interactions in the overlap region. Interestingly, dileptons can also be generated by the intense electromagnetic fields accompanying the relativistic heavy nuclei at large impact parameters~\cite{BAUR20071}, in ultra-peripheral collisions (UPC) where there is no nuclear overlap. According to the equivalent photon approximation (EPA), the electromagnetic field generated by an ultra-relativistic nucleus can be viewed as a spectrum of quasi-real photons coherently emitted by the entire nucleus~\cite{Krauss1997503} and the dilepton production process can be represented as $\gamma + \gamma \rightarrow l^{+} + l^{-}$. The equivalent two photon luminosity is proportional to $Z^{4}$, where $Z$ is the charge of the colliding nuclei. The strong dependence on $Z$ leads to copious production in relativistic heavy ion collisions. Recently, the STAR~\cite{PhysRevLett.121.132301} and ATLAS~\cite{PhysRevLett.121.212301} collaborations made measurements of dileptons at small impact parameters with nuclear overlap, and found that the electromagnetic production of dileptons can also occur in hadronic collisions~\cite{ZHA2018182,PhysRevC.97.054903}. Furthermore, a significant transverse momentum ($P_{\perp}$) broadening effect for lepton pairs produced by the two photon scattering process has been observed in non-UPCs compared to UPCs and to EPA calculations. The unsuccessful description of STAR data by STARLight model~\cite{KLEIN2017258} led to the attribution of the broadening to the possible residual magnetic field trapped in an electrically conducting QGP. Similarly, the ATLAS collaboration qualified the effect via the acoplanarity of lepton pairs in contrast to the measurement in UPC and explained the additional broadening by the multiple electromagnetic scatterings in the hot and dense medium. In Ref.~\cite{Klein:2018fmp}, S. R. Klein et al. performed calculations based on the EPA approach with different effective QED multiple scattering parameters ($\langle \hat{q}_{QED} L \rangle$) of the QGP, and described the ATLAS data well with $\langle \hat{q}_{QED} L \rangle$ of order of (50 MeV)$^{2}$ to (100 MeV)$^{2}$. These descriptions of the broadening effect assume that there is no impact parameter dependence of the transverse momentum distribution for the electromagnetic production of lepton pairs. In Ref.~\cite{ZHA2020135089}, we recover the impact parameter dependence using the lowest-order QED calculations employing the external field approximation, and found that it can describe the STAR and ATLAS data without any in-medium effect. It has also been shown that our calculations can describe the new UPC measurement from STAR~\cite{Adam:2019mby} while the STARLight calculation failed to do so.

The question remains: does the broadening effect observed by the STAR and ATLAS collaborations originate from the hot and dense medium created in relativistic heavy-ion collisions? The answer to this question crucially depends on how precisely we know the baseline transverse momentum broadening. The baseline variation on impact parameter can be precisely studied by experiment in UPCs, where there are no medium effects. However, the key experimental difficulty is the lack of a technique for the experimental selection of impact parameter ranges in UPCs. In this paper, we estimate the electromagnetic dissociation probability of nuclei for different neutron multiplicities, and demonstrate that the neutron multiplicity can be used to select different impact parameter ranges in UPCs. Employing the framework of lowest-order QED, we calculate the acoplanarity of lepton pairs from photon-photon interactions for different centralities in UPCs to probe the variation of the broadening baseline on impact parameter. These contributions provide a practical procedure for further experimental examination. 

According to the EPA method, the Coulomb excitation of an ultra-relativistic nucleus can be factorized into two parts~\cite{Baltz:1996as}: the distribution of quasi-real photons induced by the colliding nuclei, and the appropriate photon-absorption cross section of nuclei. The lowest-order probability for an excitation to the state which emits at least one neutron ($X_{n}$) is
\begin{equation}
\label{equation1}
m_{Xn}(b) = \int dk n(b,E) \sigma_{\gamma A\rightarrow A^{*}}(E),
\end{equation}
where $E$ is the photon energy, $n(b,E)$ is the flux of photons with energy $E$ at distant $b$ from the center of nucleus, and $\sigma_{\gamma A\rightarrow A^{*}}(E)$ is the photoexcitation cross section with incident energy $E$. The photon flux generated by the nucleus can be modeled by the Weizs\"acker-Williams method. For ultra-peripheral collisions, it is appropriate to employ the point-like charge distribution for the nucleus. In that case, the photon flux can be given by the simple formula~\cite{Krauss1997503}:
\begin{equation}
n(E,b) = \frac{d^{3}N}{dEd^{2}b} = \frac{Z^{2}\alpha}{\pi^{2}kE^{2}}x^{2}(K_{1}^{2}(x) + \frac{1}{\gamma^{2}}K_{0}^{2}(x)),
\label{equation2}
\end{equation}
where $\alpha$ is the electromagnetic coupling constant, $x=Eb/\gamma$, and $\gamma$ is the Lorentz factor. Here, $K_{0}$ and $K_{1}$ are modified Bessel functions of the second kind. The photoexcitation cross section $\sigma_{\gamma A\rightarrow A^{*}}(E)$ can be determined from the experimental measurements~\cite{VEYSSIERE1970561,LEPRETRE1981237,CARLOS1984573,PhysRevD.5.1640,PhysRevD.7.1362,PhysRevLett.39.737,ARMSTRONG1972445}. 

However, in high energy collisions, for example at RHIC top energy or LHC energies, $m_{Xn}(b)$ was predicted to exceed 1 at small impact parameter, which can not be interpreted as a probability. In Ref.~\cite{BROZ2020107181}, $m_{Xn}(b)$ was treated as the mean number of excitations, however, this treatment must be flawed since higher-order excitations can easily produce non-physical neutron multiplicities (exceeding the number of neutrons in the nucleus). Following Refs.~\cite{PhysRevC.64.024903,PhysRevC.60.044901}, we correlate $m_{Xn}(b)$ to the mean number of photons absorbed by the nucleus and assume a Poisson distribution for the photon multiplicity. Then the probability for absorbing zero photons, corresponding to zero neutron emission, is equal to   
\begin{equation}
P^{(0)}(b) = e^{-m_{Xn}(b)},
\label{equation3}
\end{equation}
and the probability for absorbing exactly N photons is 
\begin{equation}
P^{(N)}(b) = \frac{m^{N}_{Xn}(b)}{N!}e^{-m_{Xn}(b)}.
\label{equation4}
\end{equation}
The normalized probability density for absorbing one photon with an energy $E_{1}$ can be given by
\begin{equation}
p^{(1)}(E_{1},b) =\frac{n(E_{1},b)\sigma_{\gamma A\rightarrow A^{*}}(E_{1})}{m_{Xn}(b)}.
\label{equation5}
\end{equation}
Analogously, the probability density for absorbing $N$ photons with energies $E_{1}$, $E_{2}$,...,and $E_{N}$ is
\begin{equation}
p^{(N)}(E_{1},E_{2},...,E_{N},b) =\frac{\prod_{i=1}^{N}n(E_{i},b)\sigma_{\gamma A\rightarrow A^{*}}(E_{i})}{m^{N}_{Xn}(b)}.
\label{equation6}
\end{equation}

For a particular electromagnetic dissociation channel, i.e., $i$ neutrons emission, the probability densities for the first- and Nth-order processes can be estimated as
\begin{equation}
P^{(1)}_{i}(b) =\int dE_{1} P^{(1)}(b) p^{(1)}(E_{1},b)f_{i}(E_{1})
\label{equation7}
\end{equation}
 and
\begin{equation}
\begin{aligned}
P^{(N)}_{i}(b) =\idotsint dE_{1}...dE_{N} & P^{(N)}(b) p^{(N)}(E_{1},...,E_{N},b)\\
& \times f_{i}(E_{1},...,E_{N}).
\end{aligned}
\label{equation8}
\end{equation}
Here $f_{i}(E_{1})$ and $f_{i}(E_{1},...,E_{N})$ are the branching ratios for the considered channel of $i$ neutrons emission. We assume the simultaneous absorption of multiple photons, which gives $f_{i}(E_{1},...,E_{N}) =f_{i}(\sum_{k =1}^{N}E_{k})$. The values of $f_{i}$ for branching ratios of partial channels of different number of neutron emission are extracted from the n$\mathrm{_O^O}$n model of Ref.~\cite{BROZ2020107181}. Finally, the total probability for the emission of $i$ neutrons is $P_{in}(b) = \sum_{k=1}^{\infty} P_{i}^{(k)}(b)$. Since the probability of higher-order contributions falls down very quickly, it is enough to sum terms up to three photon absorption and neglect higher order terms.

\begin{figure}[!htb]
	\includegraphics[width=1.0\hsize]
	{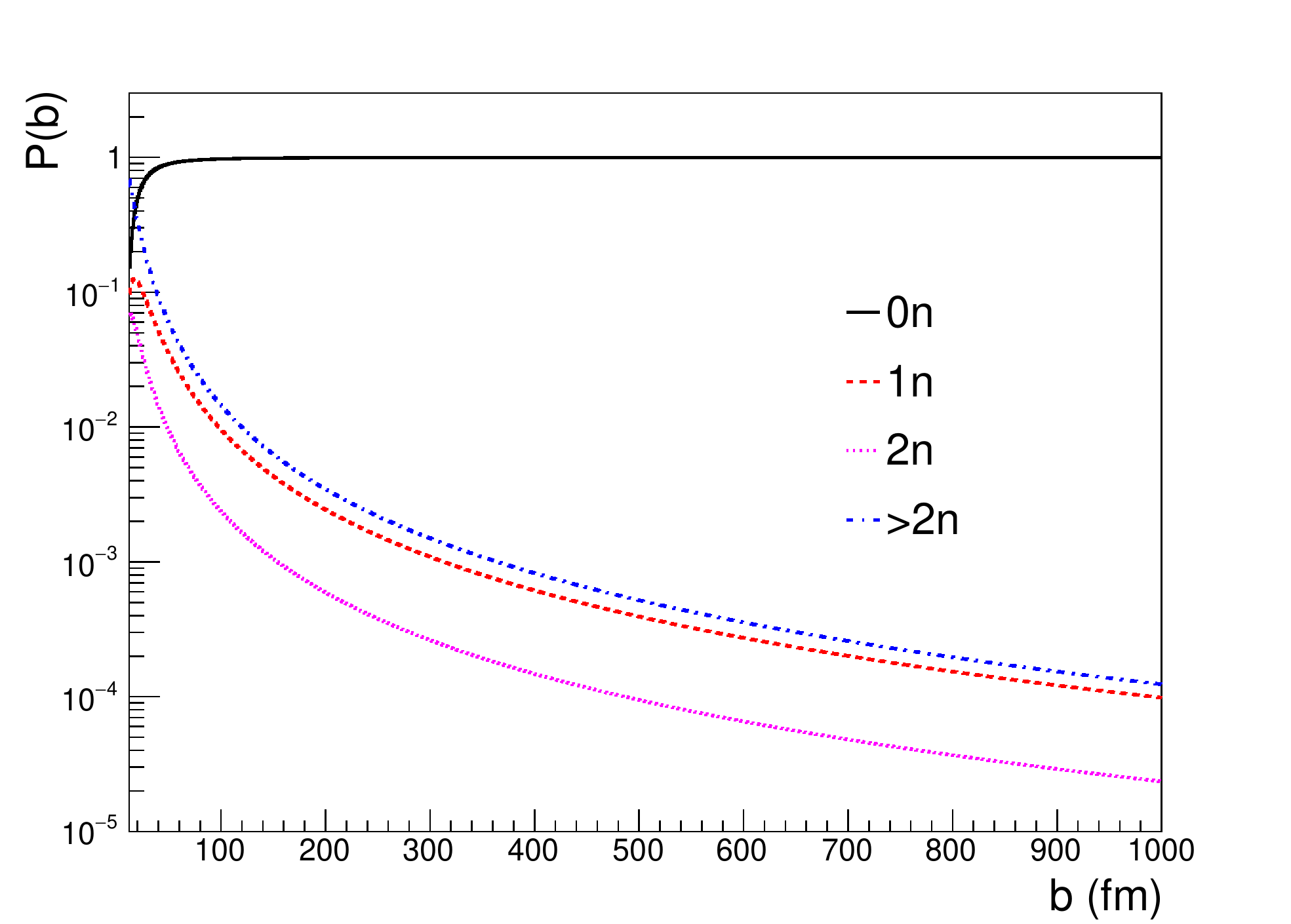}
	\caption{(Color online) Nucleus break-up probability of $^{208}$Pb as a function of impact parameter in Pb + Pb collisons at $\sqrt{s_{\rm{NN}}} = 5.02$ TeV for different number of neutron emission.}
	\label{figure1}
\end{figure}

Here we take the collisions of Pb + Pb at $\sqrt{s_{\rm{NN}}} = 5.02$ TeV as an illustration. Fig.~\ref{figure1} shows the impact parameter dependence of nuclear electromagnetic dissociation probability for different neutron multiplicities in Pb + Pb collisions at $\sqrt{s_{\rm{NN}}} = 5.02$ TeV. The dissociation probability with neutron emission drops rapidly as a function of impact parameter, which originates from the impact parameter dependence of photon flux induce by the nuclei. As shown in the figure, the dissociation process with more number of neutron emission occurs at smaller average impact parameter. This feature enables the selection of UPC events with different impact parameter ranges based on the number of neutrons detected by Zero Degree Calorimeters. The total dissociation cross section can be extracted by integrating over the impact parameter, and is found to be consistent with the experimental measurements by the ALICE collaboration~\cite{PhysRevLett.109.252302}. The probability of mutual electromagnetic dissociation, under the assumption of independent nuclear break-up, can be factorized as the product of the dissociation probabilities of each nucleus:
\begin{equation}
 P_{in_jn}(b) = P_{in}(b) \times P_{jn}(b),
\label{equation9}
\end{equation}
where the subscript $in$ and $jn$ correspond to the emission of $i$ and $j$ neutrons, respectively.

In this paper, we focus on the ultra-peripheral collisions, where the reaction should not be accompanied by hadronic interactions. According to the optical Glauber model~\cite{Miller:2007ri}, the mean number of projectile nucleons that interact at least once in A + A collisions with impact parameter $b$ is:
\begin{equation}
m_{H}(b) = \int d^{2}\vec{r} T_{A}(\vec{r} - \vec{b}) \{ 1 - exp[-\sigma_{NN}T_{A}(\vec{r})]\},
\label{equation10}
\end{equation}
where $T_{A}(\vec{r})$ is the nuclear thickness function determined from the nuclear density distribution, and $\sigma_{NN}$ is the total nucleon-nucleon cross section. Then, the probability of having no hadronic interaction is $exp[-m_{H}(b)]$.

The differential probability for the electromagnetic production of lepton pairs in heavy-ion collision at a given impact parameter, following the derivation of ~\cite{PhysRevA.51.1874,PhysRevA.55.396}, is given by
\begin{equation}
\label{equation11}
P_{l_{+}l_{-}}(b)= \int d^{2}q \frac{d^{6}P(\vec{q})}{d^{3}p_{+}d^{3}p_{-}} e^{i {\vec{q}} \cdot  {\vec{b}}},
\end{equation}
where $p_{+}$ and $p_{-}$ are the momenta of the created leptons. One can get the differential probability $\frac{d^{6}P(\vec{q})}{d^{3}p_{+}d^{3}p_{-}}$ in lowest order QED as
\begin{equation}
\label{equation12}
\begin{split}
\frac{d^{6}P(\vec{q})}{d^{3}p_{+}d^{3}p_{-}} & = (Z\alpha)^{4}
\frac{4}{\beta^{2}} \frac{1}{(2\pi)^{6}2\epsilon_{+}2\epsilon_{-}} \int d^{2}q_{1}\\
& F(N_{0})F(N_{1})F(N_{3})F(N_{4})[N_{0}N_{1}N_{3}N_{4}]^{-1} \\
& \times {\rm{Tr}}\{(\slashed{p}_{-}+m)[N_{2D}^{-1}\slashed{u}_{1} (\slashed{p}_{-} - \slashed{q}_{1} + m)\slashed{u}_{2} + \\
& N_{2X}^{-1}\slashed{u}_{2}(\slashed{q}_{1} - \slashed{p}_{+} +m)\slashed{u}_{1}] (\slashed{p}_{+}-m)[N_{5D}^{-1}\slashed{u}_{2}\\
& (\slashed{p}_{-} - \slashed{q}_{1} - \slashed{q} + m)\slashed{u}_{1} + N_{5X}^{-1} \slashed{u}_{1}(\slashed{q}_{1} + \slashed{q} - \slashed{p}_{+} \\
& + m)\slashed{u}_{2}] \},
\end{split}
\end{equation}
with
\begin{equation}
\label{equation13}
\begin{split}
& N_{0} = -q_{1}^{2},  N_{1} = -[q_{1} - (p_{+}+p_{-})]^{2},\\
& N_{3} = -(q_{1}+q)^{2}, N_{4} = -[q+(q_{1} - p_{+} - p_{-})]^{2}, \\
& N_{2D} = -(q_{1} - p_{-})^{2} + m^{2},\\
& N_{2X} = -(q_{1} - p_{+})^{2} + m^{2}, \\
&N_{5D} = -(q_{1} + q - p_{-})^{2} + m^{2},\\
& N_{5X} = -(q_{1} + q  - p_{+})^{2} + m^{2},
\end{split}
\end{equation}
where the longitudinal components of $q_{1}$ are given by $q_{10} = \frac{1}{2}[(\epsilon_{+} + \epsilon_{-}) + \beta(p_{+z}+p_{-z})]$, $q_{1z} = q_{10}/ \beta$, $\epsilon_{+}$ and $\epsilon_{-}$ are the energies of the produced leptons, and $m$ is the mass of lepton. In Eq.~\ref{equation12}, the traces and matrices have been performed using the Mathematica package Feyncalc~\cite{SHTABOVENKO2016432}. The multi-dimensional integration is performed with the MonteCarlo(MC) integration routine VEGAS~\cite{VEGAS}. As demonstrated in our previous work~\cite{ZHA2020135089}, the QED approach describes the broadening of dilepton pairs observed by the STAR and ATLAS collaborations very well in all measured centralities.  Recently, ATLAS releases new measurements with better precision and more centrality bins~\cite{ATLAS_new}, which further strengthen the validity of the approach.

Assuming that all the sub-reactions are independent, the cross section to produce a lepton pair with mutual electromagnetic excitation is 
\begin{equation}
\label{equation14}
\begin{aligned}
\sigma_{ij}(AA \rightarrow A^{*}A^{*}l_{+}l_{-})= \int d^{2}b &P_{l_{+}l_{-}}(b) \times P_{in_jn}(b)\\
&\times exp[-m_{H}(b)],
\end{aligned}
\end{equation}
where the subscript $i$ and $j$ denote the emission of $i$ and $j$ neutrons for the two colliding nuclei, respectively. Hereinafter, in the calculations, six different neutron tags are adopted to select different centrality bins in UPC: (0n,0n), (0n,1n), (0n,$>$1n), (1n,1n),(1n,$>$1n), and ($>$1n,$>$1n). Here, 0n, 1n and $>$1n represent the number of neutron emission for each nucleus, and the brackets denote the combinations of dissociation from the two beams. The centrality bins can be largely extended by choosing more combinations of neutron tags without any difficulties.

\begin{figure}[!htb]
	\includegraphics[width=1.0\hsize]
	{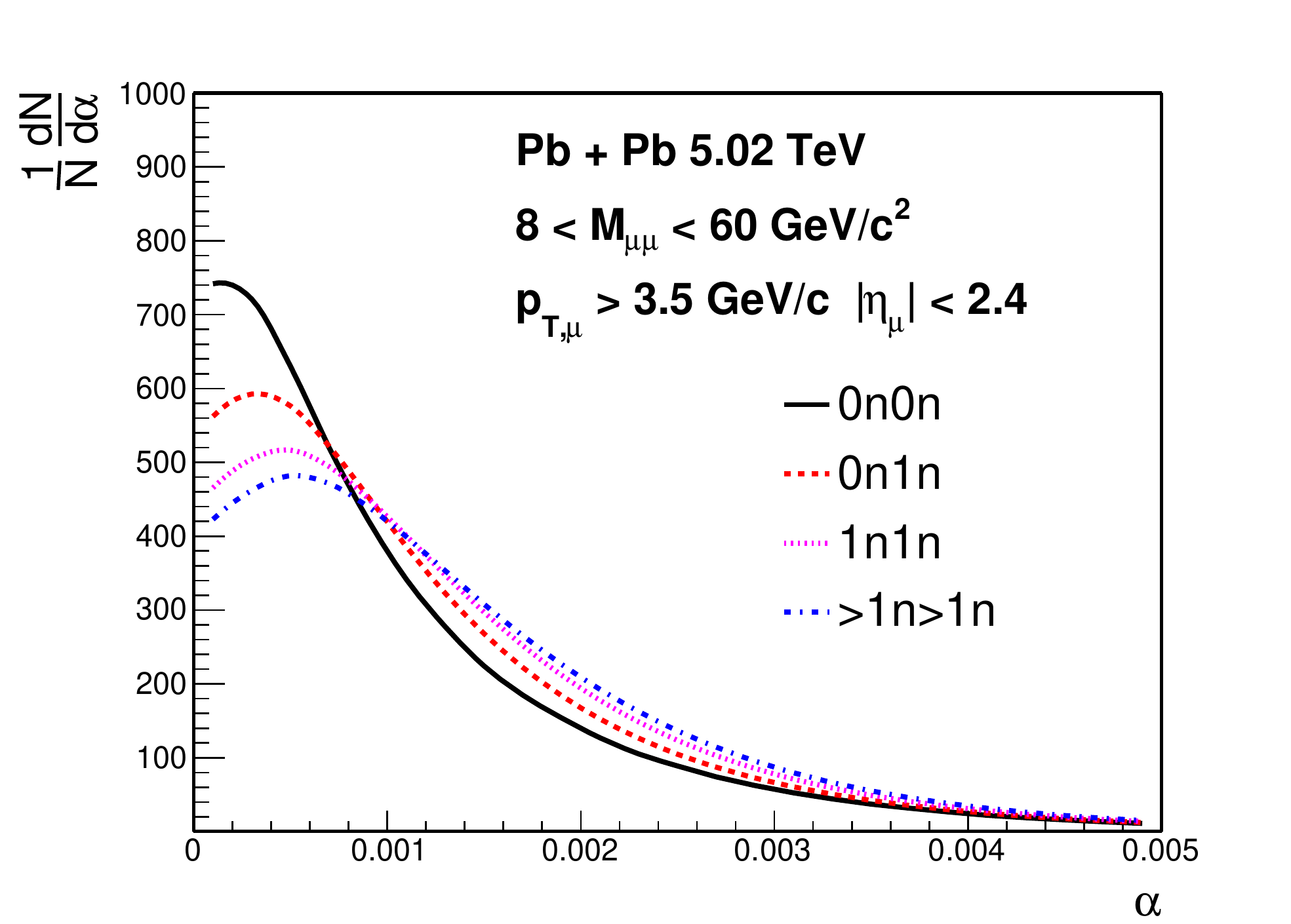}
	\caption{(Color online) The distribution of broadening variable, $\alpha$, predicted by the QED approach for muon pairs in Pb + Pb collisions at $\sqrt{s_{NN}} =$ 5.02 TeV for different centrality classes in UPCs. The results are filtered with fiducial acceptance in the text and normalized to unity to facilitate a direct comparison with experimental data.}
	\label{figure2}
\end{figure}

\begin{figure*}[!htb]
	\includegraphics[width=0.45\hsize]
	{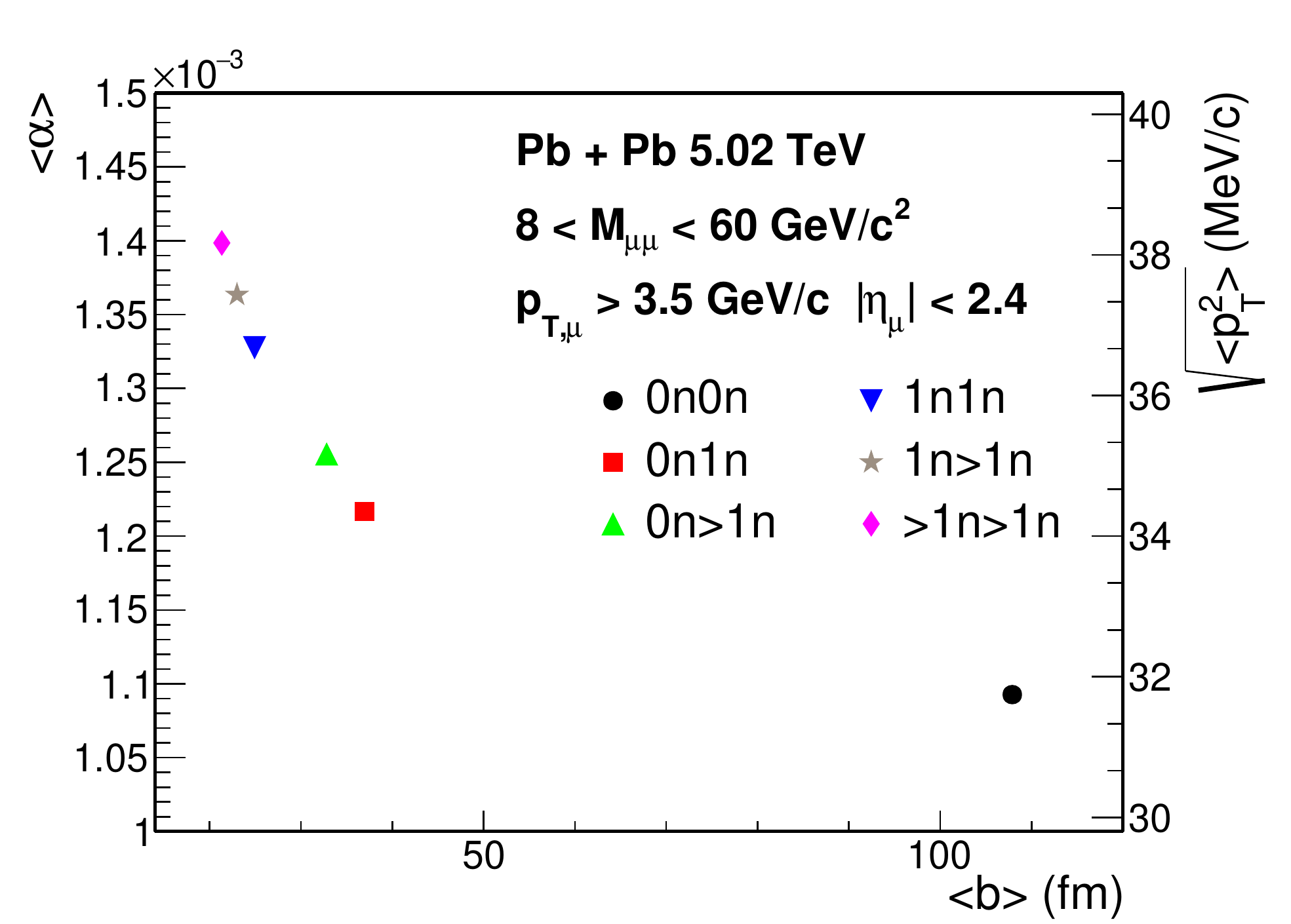}
	\includegraphics[width=0.45\hsize]
	{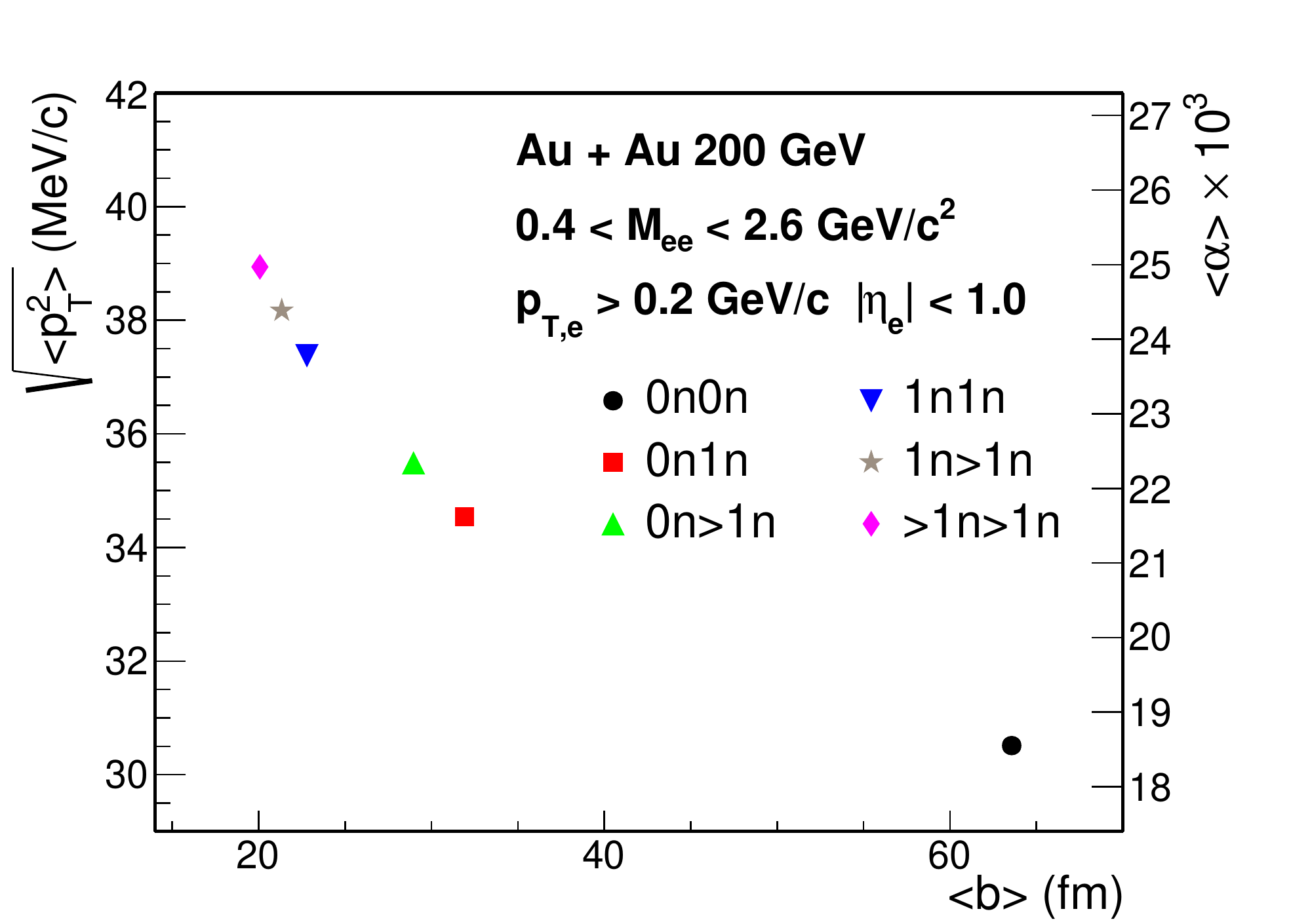}
	\caption{(Color online) Left panel: the $\langle \alpha \rangle$ and $\sqrt{\langle p_{T}^{2}\rangle}$ of muon pairs within the fiducial acceptance as a function of average impact parameter, $\langle b \rangle$, for different UPC centrality classes in Pb + Pb collisions at $\sqrt{s_{NN}} =$ 5.02 TeV. Right panel:  the $\sqrt{\langle p_{T}^{2}\rangle}$ and $\langle \alpha \rangle$ of electron-positron pairs within the fiducial acceptance as a function of average impact parameter, $\langle b \rangle$, for different UPC centrality classes in Au + Au collisions at $\sqrt{s_{NN}} =$ 200 GeV. The $\langle \alpha \rangle$ and $\sqrt{\langle p_{T}^{2}\rangle}$ is extracted for $p_{T}^{2} <$ 0.01.}
	\label{figure3}
\end{figure*}
\begin{figure*}[!htb]
	\includegraphics[width=0.45\hsize]
	{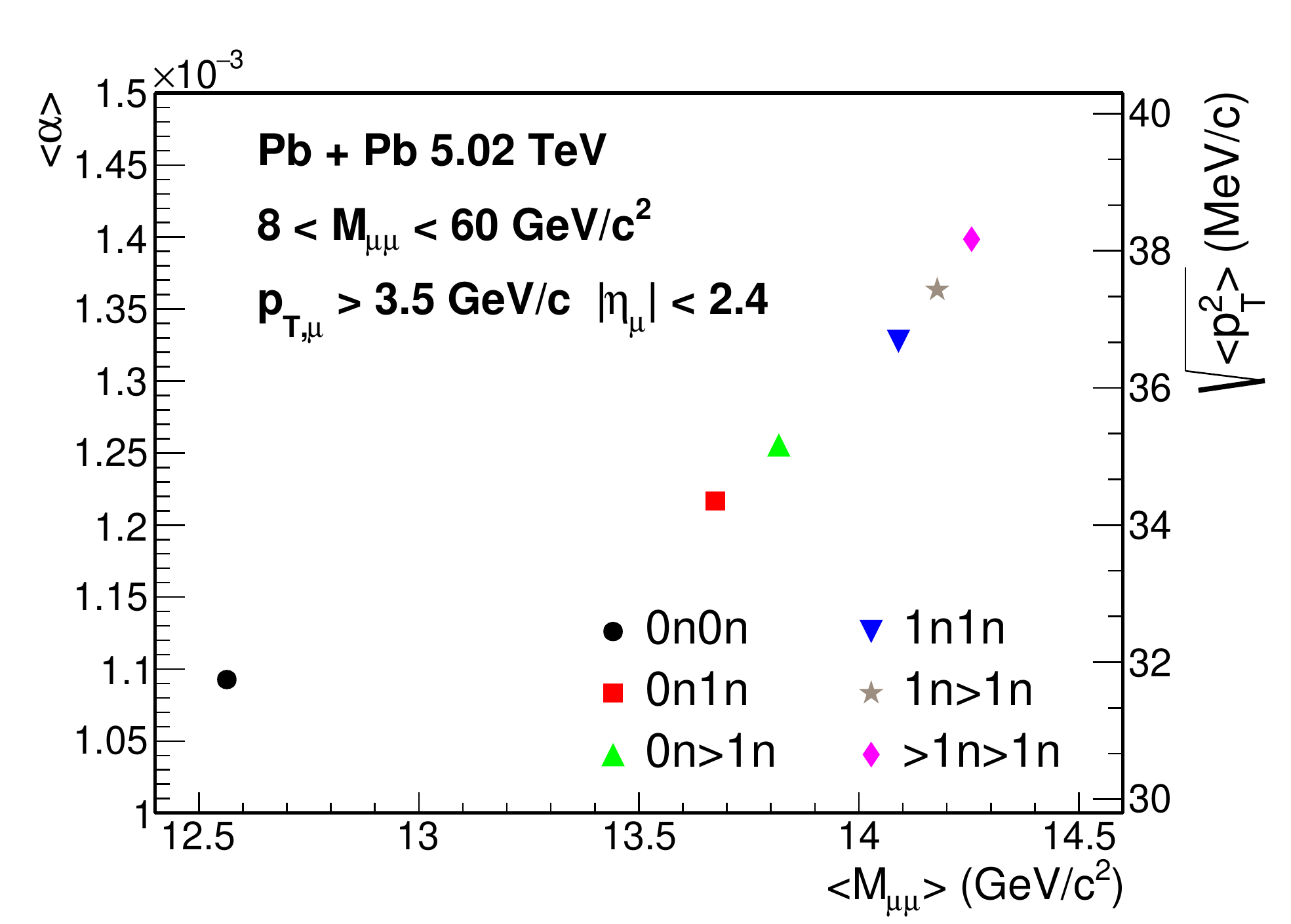}
	\includegraphics[width=0.45\hsize]
	{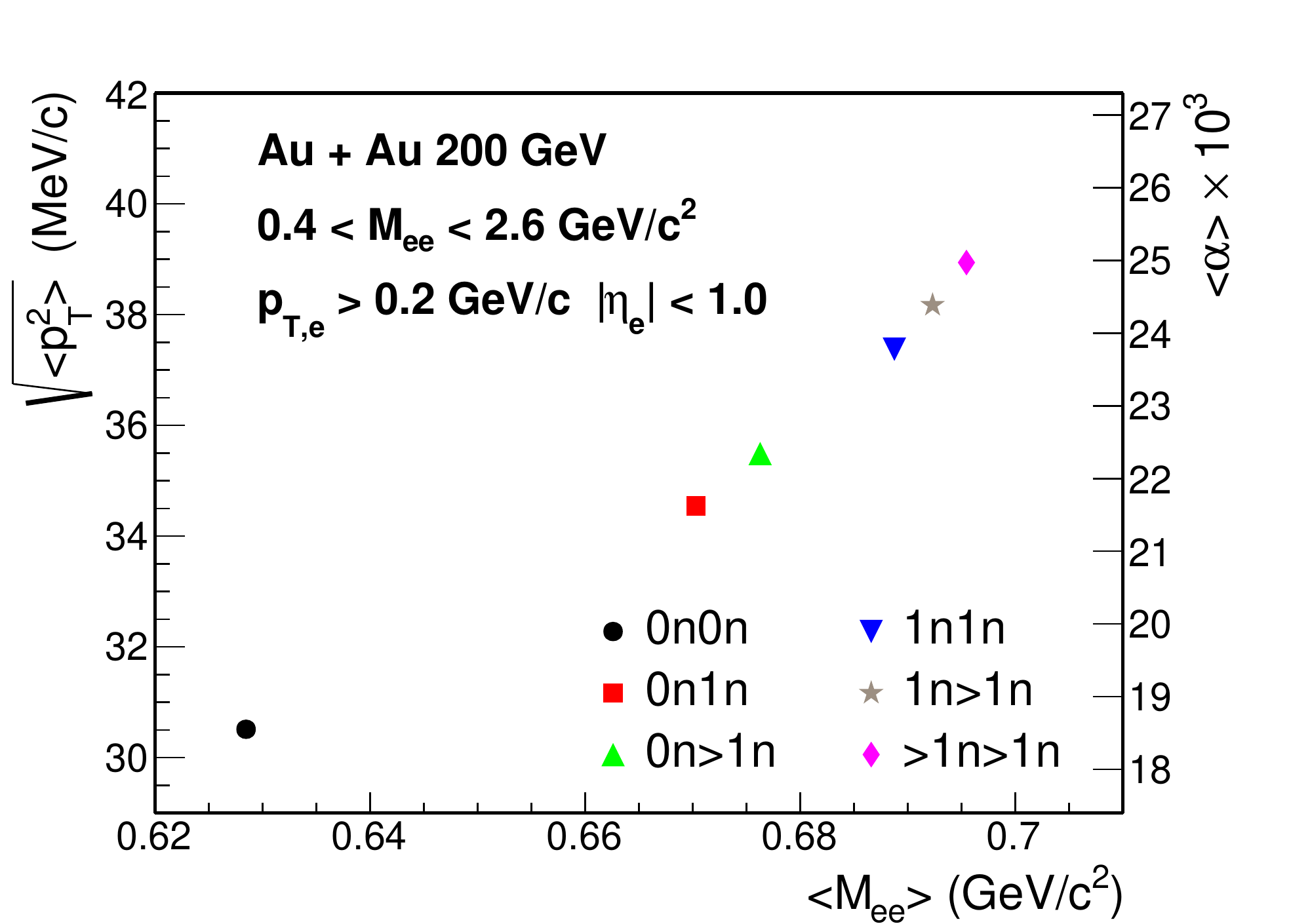}
	\caption{(Color online) Left panel: the $\langle \alpha \rangle$ and $\sqrt{\langle p_{T}^{2}\rangle}$ of muon pairs within the fiducial acceptance as a function of average pair mass, $\langle M_{\mu\mu} \rangle$,for different UPC centrality classes in Pb + Pb collisions at $\sqrt{s_{NN}} =$ 5.02 TeV. Right panel: the $\sqrt{\langle p_{T}^{2}\rangle} $ and $\langle \alpha \rangle$ of electron-positron pairs within the fiducial acceptance as a function of average pair mass, $\langle M_{ee} \rangle$,for different UPC centrality classes in Au + Au collisions at $\sqrt{s_{NN}} =$ 200 GeV. The $\langle \alpha \rangle$ and $\sqrt{\langle p_{T}^{2}\rangle}$ is extracted for $p_{T}^{2} <$ 0.01.}
	\label{figure4}
\end{figure*}

At RHIC, the broadening is directly extracted from the transverse momentum spectra of electron-positron pairs. While at LHC, the broadening effect is characterized by the acoplanarity correlations of lepton pair, which takes advantage of better angular measurement than momentum measurement of high momentum leptons. The pair acoplanarity $\alpha$ is defined as:
\begin{equation}
\label{equation15}
\alpha = 1 - \frac{|\phi^{+} - \phi^{-}|}{\pi},
\end{equation}
where $\phi^{\pm}$ the azimuthal angles of the two individual leptons. This definitions largely avoids the detector induced distortions from poor momentum resolution. Fig.~\ref{figure2} shows the results of our calculations for $\alpha$ distributions of muon pairs in Pb + Pb collisions at $\sqrt{s_{NN}} =$ 5.02 TeV for different neutron emission scenarios in UPCs. The results are filtered with the fiducial acceptance described in the figure and normalized to unity to facilitate a direct comparison with experimental data. The $\alpha$ distribution with no neutron emission from the two nuclei (labelled as ``0n0n'' in the figure) have a narrower distribution comparison to the same distribution for events with any number of neutron emission. As expected, the normalized $\alpha$ spectrum becomes broader in the case of emitting more neutrons, which correspond to smaller impact parameters. Interestingly, the most probable value of $\alpha$ distribution is not at zero, and shifts to a higher value in the collisions with more neutron emission. As pointed out in Ref.~\cite{Klein:2018fmp}, the high-order soft photon radiation will modify the $\alpha$ distribution, which may complicate these calculations. Fortunately, according to the Sudakov resummation approach, the effect is small at low pair $P_{\perp}$ (large $\alpha$) . In this paper we focus on the  $p_{T}^{2} <$ 0.01 range which accounts for the majority of the production and neglect the Sudakov effect, which should be studied in future work.

To quantitatively describe the impact parameter dependent broadening for lepton pair production in UPCs, we employ the QED approach to estimate the mean of the $\alpha$ and $\sqrt{p_{T}^{2}}$ distributions ($\langle \alpha \rangle$, and  $\sqrt{\langle p_{T}^{2}\rangle}$) versus average impact parameter, $\langle b \rangle$, for different neutron emission scenarios. Fig.~\ref{figure3} shows the $\langle \alpha \rangle$ and  $\sqrt{\langle p_{T}^{2}\rangle}$ of lepton pairs as a function of $\langle b \rangle$ for different neutron emission scenarios. The results are filtered with the fiducial acceptance described in the figure, and the $p_{T}^{2}$ range is limited to $<$ 0.01 to avoid the influence of the Sudakov effect, which is not taken into account in the calculations. As shown in the figure, the average impact parameter of lepton pair production, $\langle b \rangle$, varies significantly for different combinations of neutron multiplicity from the two colliding nuclei. The $\langle \alpha \rangle$ and  $\sqrt{\langle p_{T}^{2}\rangle}$ decrease with increasing $\langle b \rangle$, which reflects more broadening in UPCs with smaller $\langle b \rangle$. The impact parameter dependence is much stronger at small $\langle b \rangle$  than that at large $\langle b \rangle$. The $\langle \alpha \rangle$ and  $\sqrt{\langle p_{T}^{2}\rangle}$ for ($>$1n$>$1n) are 20$\sim$ 30\% larger than that for (0n0n), which is signifiant for further experimental test.

In experiment, it is impossible to determine the impact parameter ordering for different neutron emission scenarios in UPC from direct measurements. According to the QED and EPA calculations, the pair mass distributions of lepton pairs vary with impact parameter, which could be used to denote the impact parameter ordering in UPCs. With smaller impact parameter, there will be a higher probability for the nucleus to emit energetic photons, which gives more production of lepton pairs with high pair mass. Fig.~\ref{figure4} shows the $\langle \alpha \rangle$ and  $\sqrt{\langle p_{T}^{2}\rangle}$ of lepton pairs within the fiducial acceptance as a function of average pair mass, $\langle M_{l^{+}l^{-}} \rangle$, for UPCs with different neutron emission scenarios. As depicted in the figure, the $\langle M_{l^{+}l^{-}} \rangle$ can be considerably altered by selecting different impact parameter ranges in UPCs. The $\langle \alpha \rangle$ and  $\sqrt{\langle p_{T}^{2}\rangle}$ increase with increasing $\langle M_{l^{+}l^{-}} \rangle$, which also reveals more broadening at smaller impact parameter.

In summary, we calculate transverse momentum broadening for the electromagnetic production of dileptons in ultra-peripheral heavy ion collisions accompanied by mutual nuclear dissociation. The mutual nuclear dissociation with different neutron multiplicities is an effective tag to determine the impact parameter range in UPCs. The broadening effects show significant dependence on the number of emitted neutrons in UPCs, which calls for further experimental validation. The verification of the impact parameter dependence in UPCs would provide a solid baseline to exact the possible medium effects in hadronic heavy-ion collisions.

  This work was funded by the National Natural Science Foundation of China under Grant Nos. 11775213, 11505180 and 11375172, the U.S. DOE Office of Science under contract No. DE-SC0012704, and MOST under Grant No. 2014CB845400.

\nocite{*}
\bibliographystyle{aipnum4-1}
\bibliography{aps}
\end{document}